\documentclass[reprint,amssymb, amsmath, aip,cha,twocolumn]{revtex4-1}
\usepackage{graphicx}
\usepackage[caption = false]{subfig}
\usepackage{color}
\usepackage{epsfig}
\usepackage{bm}%
\usepackage[colorlinks=true,linkcolor=blue]{hyperref}%
\expandafter\ifx\csname package@font\endcsname\relax\else
 \expandafter\expandafter
 \expandafter\usepackage
 \expandafter\expandafter
 \expandafter{\csname package@font\endcsname}%
\fi
\begin{document}
\title{Flowing dusty plasma experiments: Generation of flow and measurement techniques}%
\author{S. Jaiswal}%
\email{surabhijaiswal73@gmail.com}
\author{P. Bandyopadhyay}
\author{A. Sen}
\affiliation{Institute For Plasma Research, Bhat, Gandhinagar,Gujarat, India, 382428}%
\date{\today}
%**************************************************************
%#####################################################################################                  ABSTRACT
%************************************************************************************************
\begin{abstract}
A variety of experimental techniques for the generation of subsonic/supersonic dust fluid flows and means of  measuring such flow velocities are presented. The experiments have been carried out in a $\Pi-$shaped Dusty Plasma Experimental (DPEx) device with micron size kaolin/Melamine Formaldehyde (MF) particles embedded in a background of Argon plasma created by a direct current (DC) glow discharge. A stationary dust cloud is formed over the cathode region by precisely balancing the pumping speed and gas flow rate. A flow of dust particles/fluid is generated by additional gas injection from a single or dual locations or by altering the dust confining potential. The flow velocity is then estimated by three different techniques, namely, by super Particle Identification (sPIT) code, Particle Image Velocimetry (PIV) analysis and the excitation of Dust Acoustic Waves (DAWs). The results obtained from these three different techniques along with their merits and demerits are discussed. An estimation of the neutral drag force responsible for the generation as well as the attenuation of the dust fluid flow is made. These techniques can be usefully employed in laboratory devices to investigate linear and non-linear collective excitations in a flowing dusty plasma.  
\end{abstract}

\maketitle
\section{Introduction}
During the last couple of decades, a great deal of research has been devoted to the new and fascinating field of dusty (complex) plasmas. A complex or dusty plasma consists of the usual two component plasma supplemented by the addition of micron or sub-micron sized dust grains \cite{shukla2009,morfill2009}. In a low-temperature  dusty plasma, highly mobile electrons reside on the surface of these heavy dust particles and make the grains negatively charged \cite{barken1994}. Unlike the other conventional plasma species, these highly charged dust particles interact strongly with each other and can even form ordered structures like a crystal \cite{ikezi1986,thomas1994}. A dusty plasma medium also supports a rich variety of collective phenomena$-$linear/nonlinear waves and instabilities \cite{rao1990,barken1995,pintu2008, nakamura2012, heinrich2009} due to the additional degree of freedom provided by the charged dust particles. Dust and dusty plasmas are ubiquitous in nature  \cite{shukla2000} e.g., in planetary rings, comet tails, and interplanetary and interstellar clouds. Dust particles are often present in plasmas used for industrial applications \cite{merlino2006} and in thermonuclear fusion devices like tokamaks \cite{nakamura2006}. \par
Flow induced excitation of linear and non-linear waves is an emerging area of research in the field of dusty/complex plasmas. In recent times, a series of experimental studies \cite{nakamura2012, heinrich2009, fink2013, usachev2014, samsonov2003} have been carried out world wide on the excitation of non-linear waves in a flowing dusty plasma. In 2003, Samsonov \textit{et al.}\cite{samsonov2003} reported the experimental observation of shock structures in a rf produced 3D complex plasma under microgravity conditions in the PKE-Nefedov device by applying a sudden gas pulse using an electromagnetic valve. Later, Nakamura \cite{nakamura2012} experimentally observed a bow shock like formation in a 2D flowing dusty plasma. Very recently, Fink \textit{et al.} \cite{fink2013} triggered auto waves in a complex plasma by injecting gas from a gas-flow controller. In most of these  experiments the flow of dust fluid was initiated either by tilting the complete experimental set-up \cite{nakamura2012} to use gravity or by using gas puffs \cite{ fink2013, samsonov2003}. Tilting the device can often be problematic as it can disturb optical alignments. In addition, it also limits the amount of control one has on the flow of particles. Likewise experiments with gas puffs also have limitations in the way of induced changes in the equilibrium configuration due to the sudden introduction of neutral gas into the device and the concommitant increase in the pressure.\par
After generating the flow in the dust fluid, it is equally important to accurately measure the velocity of the flow so as to properly decipher the underlying physical mechanism responsible for the flow induced excitations of linear/nonlinear waves and vortex structures. To study the properties of these waves/structures in a detailed manner, it also sometimes becomes essential to make a coordinate transformation from the laboratory frame to the fluid frame for which it is necessary to know the fluid velocity. Furthermore, several theoretical studies predict that the wave behaviour undergoes drastic changes when the fluid flow velocity changes its magnitude from subsonic to supersonic values. In some of the dusty plasma experiments, in which the dust dynamics plays an important role, the estimation of different forces (e.g. neutral drag force, ion drag force, electrostatic force etc.) acting on the flowing dust particles are very crucial. For such experiments, time varying measurements of the fluid velocity can provide useful information to estimate these forces. Thus precise measurements of the fluid velocity are crucial  for investigating the propagation characteristics of flow induced excitations of linear/non-linear waves as well as for estimating the fundamental forces acting on the dust particles.  \par
Measurements of flow velocities in a dusty plasma remain a relatively unexplored area of experimental research and there is a need to identify, develop and document techniques for carrying out such measurements. In this article, we report on different experimental techniques that we have implemented to initiate dust flow in a tabletop Dusty Plasma Experimental (DPEx) device and also describe the variety of tools that we have employed to measure the flow velocities. Among these measurement techniques, to the best of our knowledge, the use of low amplitude Dust Acoustic Waves has not been tried before and our findings should stimulate further research and development in this area. This novel technique allows us to measure the velocity of a dense fluid which flows with a very high velocity. The results obtained by these different techniques show a good agreement with each other. A detailed discussion on the merits and demerits of the different techniques to initiate the flow and of the various measurement tools is presented. An estimation of the neutral drag force, which acts to either induce fluid flow or slow it down, is also provided. 
%Few proposed experiments on flow induced excitation of linear/nonlinear waves and structures in DPEx device are %listed at the end. 
\par
The paper is organised as follows: in the next section (Sec.~\ref{expt}), we describe the Dusty Plasma Experimental (DPEx) device, the plasma production method and the associated diagnostics for measuring plasma parameters. Sec.~\ref{flow} describes the different techniques that we have followed to generate the flow in a dusty plasma. The measurement of fluid velocity using different tools with their merits and demerits are presented and discussed in Sec.~\ref{tech}. The main results obtained in the experiments are summarised in Sec.~\ref{result} and a few concluding remarks are made in Sec.~\ref{conclusion}.
%%%%%%%%%%%%%%%%%%%%%%%%%%%%%%%%%%%
\section{Experimental device and procedure \label{expt}}
\begin{figure}[ht]
%\centering{\includegraphics[scale=0.45]{../../figures/chamber_iop_2}}
\centering{\includegraphics[scale=0.45]{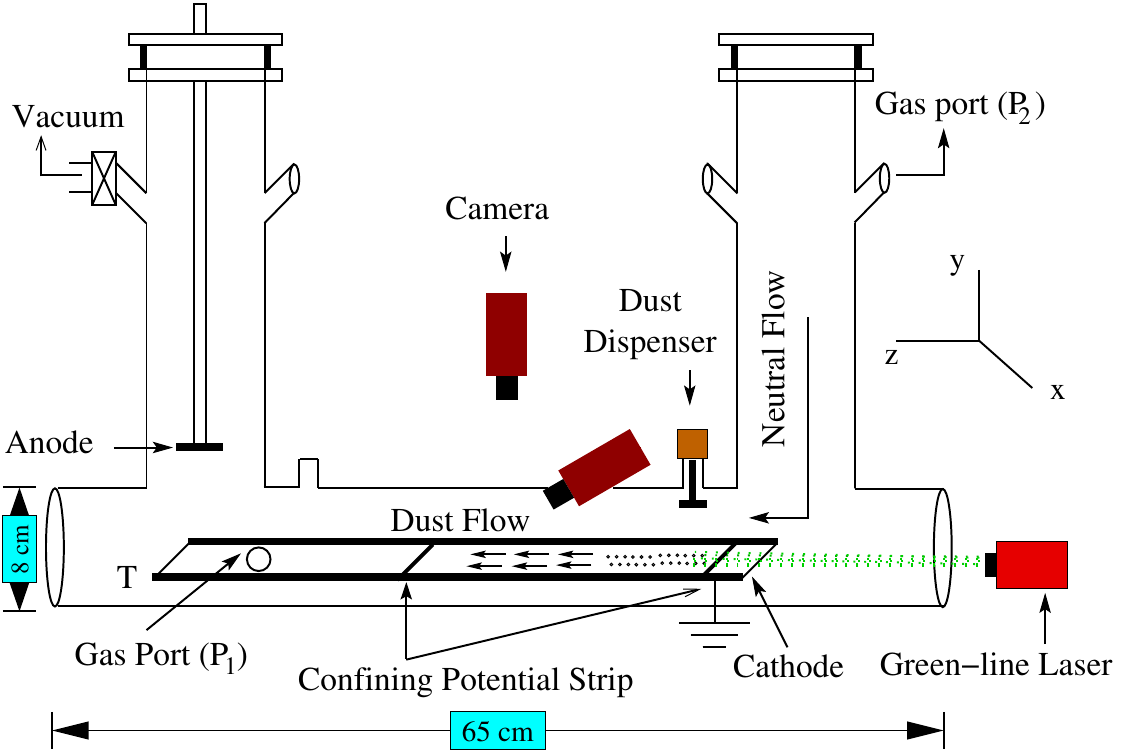}}
\caption{(a) Schematic diagram of Dusty Plasma Experimental (DPEx) setup., T: grounded cathode {tray}. }
\label{fig:fig1} 
\end{figure}

The experiments are performed in a {$\Pi-$} shaped Dusty Plasma Experimental (DPEx) device which is made of pyrex glass.  A schematic diagram of DPEx device along with its associated diagnostics is shown in Fig.~\ref{fig:fig1}. The system geometry is inspired by that of the PK-4 \cite{pk42004} device. However there are essential differences between these two devices which has been reported in detail elsewhere \cite{jaiswal2015}. A rotary pump is used to evacuate the system up to its base pressure of $10^{-3}$~mbar and Argon gas is introduced by a mass flow controller and gas dosing valve through the gas ports, P$_1$ and P$_2$. A disc shaped anode of 3 cm diameter and a long grounded cathode tray of 2 mms thickness, 6.1 cms width and 40 cms length placed inside the connecting tube, are used as electrodes for plasma generation. The edges of the cathode tray provide the radial confinement whereas a couple of stainless steel strips on the cathode provide the axial confinement to the dust cloud. Depending upon the experimental requirement, the kaolin dust particles are either sprinkled on the cathode or MF particles are injected into the plasma by using the dust dispenser. 
 %%%%%%%%%%%%%%%%%%%%%%%%%  
%%%%%%%%%%%%%%%%%%%%%%%%%  
To operate the device for flowing dusty plasma experiments, we begin with the pumping of the experimental set-up. After achieving the required base pressure of $\sim$ 0.001 mbar, argon gas is flushed several times and pumped down to its base pressure. Finally, the working pressure is set in the range of $0.1-0.2$~mbar by maintaining the pumping speed and the gas flow rate. A DC glow discharge plasma is formed in between the anode and cathode by applying a discharge voltage in the range of $300-400$~volt. In some of the experiments, the MF dust particles are introduced into the plasma by shaking the dust dispenser. In other experiments, kaolin particles are spread uniformly on the cathode before closing the device. These MF or Kaolin particles get negatively charged by collecting more electrons than ions and get trapped in the plasma sheath boundary above the grounded cathode. In this levitated condition, the vertical component of the sheath electric field provides the necessary electrostatic force to the particles to balance the gravitational force. The radial and the axial sheath electric fields are responsible for the radial and axial confinement of the dust particles against their mutual Coulomb repulsive forces. By adjusting the pumping speed and the gas flow rate, a steady-state equilibrium dust cloud can be formed over the cathode in between the two stainless steel strips. The particle cloud is illuminated by a green line-laser light (532 nm, 100 mW) along the axial length of the connecting tube so that the dust cloud can be seen over its entire length. The Mie-scattered light from the dust particles is captured by a couple of CCD cameras and the images are stored into a high speed computer. The high speed camera (60fps, 1MP) is placed at an angle of ${15^\circ}$ with the y-axis and the high resolution camera (15fps, 4MP) is placed exactly perpendicular to the dust cloud. The speed of the cameras can be further increased at the cost of lowering their resolutions.
%%%%%%%%%%%%%%%%
\section{Flow generation \label{flow}}
In the present set of experiments, the flow of dust particles is initiated after attaining a steady state equilibrium dust cloud. In the subsequent sections, we will discuss about the three different techniques that we have employed to generate the flow of dust particles. 
\subsection{By single gas injection \label{single}}
As mentioned above, an equilibrium dust cloud can be formed when the pumping rate (20\% opening) and the gas flow rate (27.5 ml$_s$/min) through gas  port P$_1$ are  balanced in a precise manner. If the pumping rate exceeds the gas flow rate, the particles are seen to flow from right to left and in the reverse direction if the gas feeding rate is increased beyond 27.5 ml$_s$/min. However, for our experimental convenience, the flow is always generated from right to left by reducing the gas flow rate in steps of 2.75 ml$_s$/min. from its equilibrium value. After initiating the flow of dust particles, the gas feed rate is set back to its original value within a second and hence the equilibrium condition is restored. As a result almost all the particles are found to return to their original position from where they had started their journey. In this process some particles always get lost during the experiments. The particle velocity in this procedure can be raised up to $20-25$ cm/sec by increasing the flow rate difference. When the flow rate difference is set to a higher value, the particles move towards the pump with very high velocity and hence they overcome the potential barrier and flow over the strip and finally fall down on the left edge of the glass tube where the cathode ends.
\subsection{By dual gas injection \label{dual}}
In an alternative method, the flow of dust particles is initiated by using two gas feed ports $P_1$ and $P_2$ as shown in Fig.~\ref{fig:fig1}.  As discussed, initially the steady state dust cloud is formed at a particular discharge condition, in between two confining potential strips by adjusting the pumping rate and the gas feeding rate through port $P_1$. In this equilibrium condition, the dust particles only show Brownian motion due to their thermal energy. It is to be noted that  the gas dosing valve mounted at gas feeding port $P_2$ is kept closed at this point. To generate the flow of dust particles, this gas dosing valve is opened and as a result the particles are found to flow from right to left. In this dual gas injection technique, the flow velocity can be changed very precisely by changing the gas flow rate of port $P_2$. With the help of this method we can maintain a constant gas pressure while performing the experiments. In some of the experiments at lower gas pressure (below $P=0.09$~mbar), we can produce a flow in the particles by first attaining the background pressure at equilibration by using both the ports and subsequently injecting the dust particles by the dust dispenser.
%%%%%%%%%%%%%%%%%%%%%%%%
\begin{figure}[!hb]
%\centering{\includegraphics[scale=0.8]{../../figures/fig2mmm}}
\centering{\includegraphics[scale=0.8]{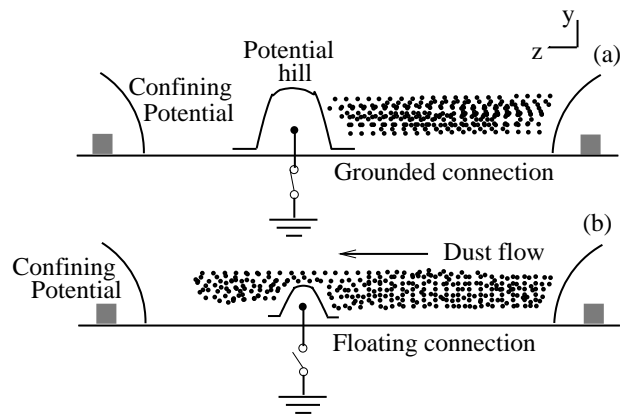}}
\caption{(a) Equilibrium dust configuration with the potential hill created by a grounded wire and b) dust flow induced by sudden lowering of the potential hill by switching the grounded connection to a floating connection.}
\label{fig:potential1} 
\end{figure}
%%%%%%%%%%%%%%%%%%%%%%%
\subsection{By altering the confining potential \label{confinement}}
In the third technique of flow generation, the equilibrium of the long dust cloud is first achieved in between the confining potential strips (located at right end of the cloud) and the grounded wire (located at left end of the cloud) as shown schematically in Fig.~\ref{fig:potential1}(a). There is an experimental arrangement by which this wire can be kept either at floating or at ground or at intermediate potentials. The grounded wire creates an electrostatic potential hill by which the dust particles get confined in the axial direction. To initiate the flow, the grounded wire is suddenly switched to the floating potential (or into a potential which has a higher value than the ground) and as a result the stationary dust cloud is found to flow over the wire. Fig.~\ref{fig:potential1}(b) shows a schematic diagram of this situation when the height of the potential hill is reduced by changing the wire potential from ground to floating. In this condition, the particles cannot feel the presence of the wire and are observed to flow over it. The height of the potential hill and hence the speed of flow of the particles, can be precisely controlled by drawing currents through different combinations of resistances connected between the wire and the ground. It is worth while mentioning that, in this scheme of flow generation the neutrals do not carry the dust particles. \par
Among these three different techniques, the single and dual gas injection techniques are used to generate a motion of dust particles by inducing a neutral gas flow in the experimental device. In other words, the dust particles flow from right to left due to the neutral streaming which carry the dust particles along their way. With the help of these schemes, the dust fluid velocity can be raised from subsonic (few mm/sec to cm/sec) to supersonic (few tens of cm/sec) values.  However, the single injection technique creates a momentary (for a time less than a second) change of the neutral gas pressure which disturbs the steady state equilibrium  during the course of the experiment. This problem is overcome in the second technique of dual gas injection by introducing the neutral gas in a continuous manner to maintain a constant gas pressure during the experiments. The third technique, namely flow generation by altering the confining potential is very useful for experiments which demand a stationary neutral gas. However this technique is limited by its inability to produce very high velocity fluid flows for long periods in comparison to the first two techniques. 

\section{Techniques of flow velocity measurements \label{tech}}
After the initiation of the fluid flow it is essential to measure the dust fluid velocity prior to conducting any further experiments. The fluid velocity measurement helps in estimatimg the  fundamental forces that act on the particles and that influence the collective behaviour of the dusty plasma. The velocity measurement is also needed to determine the true velocity of the linear/nonlinear wave structure from the laboratory frame measurements. We now discuss various techniques of measuring the fluid flow velocity in the following subsections.
\subsection{By super Particle Identification Tracking (sPIT) code \label{spit}}
We begin with an Idl based super Particle Identification Tracking (sPIT) \cite{konopka2000, goree2007} code to measure the dust particle velocity from an analysis of video images of the flow. This code efficiently measures the flow velocity when the individual particles are distinguishable in consecutive video frames. Hence, for an accurate measurement of the particle velocity using the sPIT code, one needs to restrict oneself to conducting experiments where the dust number density is low (not more than 30 particles in the field of view of $9.0$ mm$\times 2.2$ mm) and that move with a moderate velocity (not more than $\sim 8$ cm/sec). This is to ensure that we focus on a small area at a maximum frame rate so that the particles can be detected very easily. The analysis shows that the particles  start to move with zero velocity and then accelerate to a certain value and finally attain a terminal velocity within this region.
%%%%%%%%%%%%%%%%%%%%%%%%
\begin{figure}[!ht]
%\centering{\includegraphics[scale=0.8]{../../figures/color_plot_spit_code_m1}}
%\centering{\includegraphics[scale=0.4]{../../figures/fig3_m}}
\centering{\includegraphics[scale=0.4]{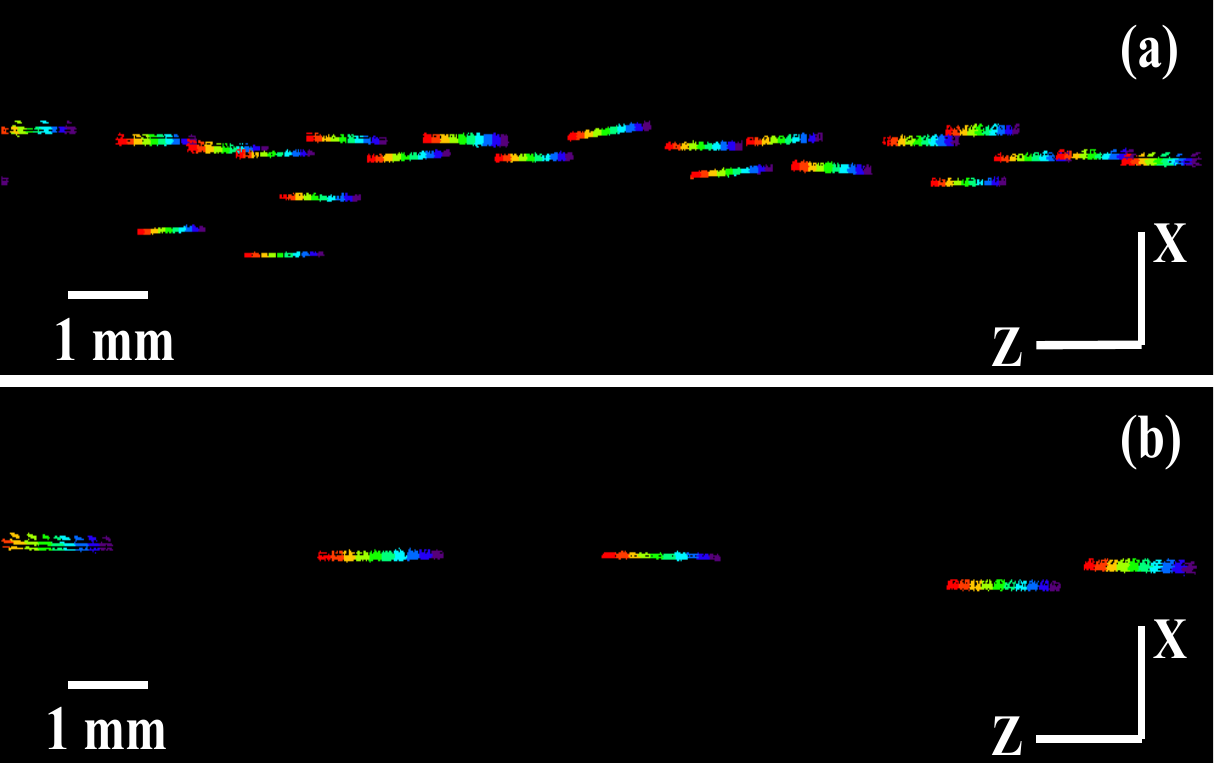}}
\caption{ Color plot showing the particle trajectory at (a) 0.15 mbar and  b) 0.13 mbar respectively at discharge voltage (V$_d$) of 300 volt .}
\label{fig:spit} 
\end{figure}
%%%%%%%%%%%%%%%%%%%%%%%% 
 The proper threshold and background values are set for the consecutive images to trace the particle with maximum probability for all the frames. Fig.~\ref{fig:spit} depicts the trajectory of the particles for two different pressures at a discharge voltage of 300 Volt. This figure is created by overlapping ten consecutive frames that are marked in different colors in the color sequence of a rainbow. The violet colour corresponds to the positions of the particles in the first frame whereas the red colour corresponds to their positions in the tenth frame. From such a sequence it is clear that the particles are moving from right to left. Fig.~\ref{fig:spit}(a) represents the particle trajectory for 0.15 mbar whereas Fig.~\ref{fig:spit}(b) represents the same for 0.13 mbar pressure. It is clearly seen in the figure that the average particle trajectory becomes shorter at higher pressures. It essentially  indicates that the particles undergo a larger number of collisions with the neutrals at P $=0.15$~mbar and as a result they are not able to travel a longer distance. It is also to be noted that by calculating the distance traveled in 10 frames, we are able to estimate the velocity of the particles.\par
%%%%%%%%%%%%%%%%%%%%%%%%
\begin{figure}[!ht]
%\centering{\includegraphics[scale=0.8]{../../figures/velocity_vs_distance2_spit_code_m}}
%\centering{\includegraphics[scale=0.8]{../../figures/spitvelocity_Profile1}}
\centering{\includegraphics[scale=0.8]{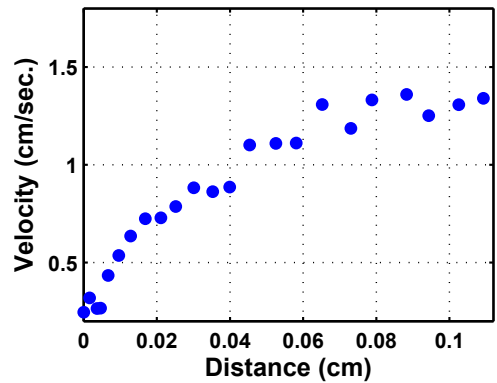}}
\caption{A plot showing the variation of velocity with the distance travelled by the particles at P $=0.13$ mbar, discharge voltage (V$_d)=310$ volt and for the flow rate difference of 2.75 ml$_s$/min.}
\label{fig:spitVelocity} 
\end{figure}
%%%%%%%%%%%%%%%%%%%%%%%% 
The spatial variation of the velocity of an individual particle is depicted in Fig.~\ref{fig:spitVelocity}. It shows that the particle experiences an acceleration in the beginning and then it attains a terminal velocity which is approximately equal to 1.3~cm/sec. The particles achieve the terminal velocity due to the neutral drag force which always opposes the motion of the dust particles when they travel faster than the neutrals.  \\    
\subsection{By Particle Image Velocimetry (PIV) analysis \label{piv}}
\begin{figure}[!ht]
%\centering{\includegraphics[scale=0.75]{../../figures/vector_plot_m6_1}}
\centering{\includegraphics[scale=0.75]{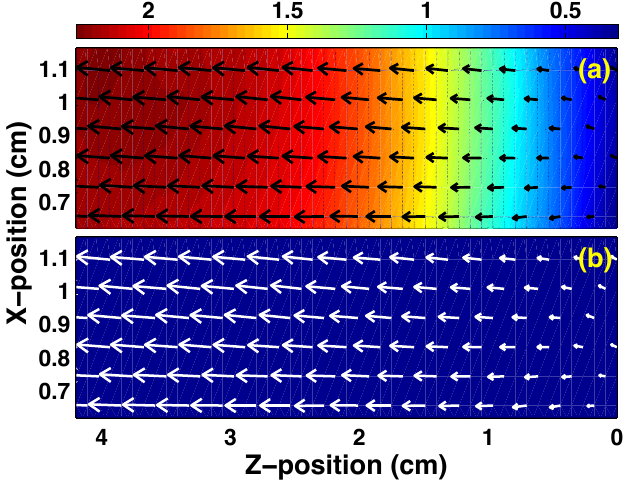}}
\caption{velocity vector fields along with the magnitude of the velocity components $v_z$ (Fig.~\ref{fig:PIV1}(a)) and $v_x$ (Fig.~\ref{fig:PIV1}(b)) respectively.}
\label{fig:PIV1} 
\end{figure}

\begin{figure}[!ht]
%\centering{\includegraphics[scale=0.8]{../../figures/velocity_profile_piv_m}}
\centering{\includegraphics[scale=0.8]{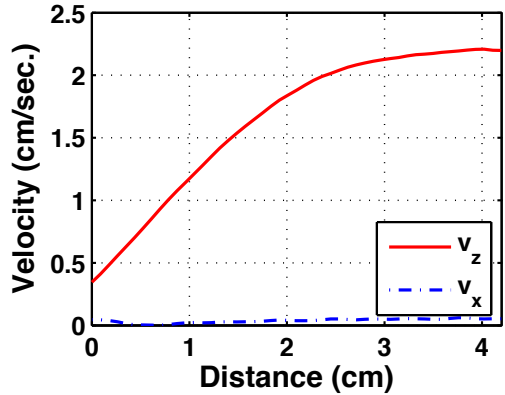}}
\caption{The variation of velocity component $v_z$ (shown by solid line) and $v_x$ (shown by dashed line) with distance travelled by the particles.  }
\label{fig:vxvy} 
\end{figure}

In addition to the sPIT code, a Matlab based open access Particle Image Velocimetry (PIV) \cite{PIV2014} analysis has also been carried out to estimate the average velocity of the particles. For the present PIV analysis, 50 still frames of pixel resolution $1000 \times 225$ in the interval of 30.7 m.sec of flowing dust particles are considered.  A 2-pass algorithm is used in which $64 \times 64$ sq. pixel interrogation area in steps of 32 pixel followed by $32 \times 32$ sq. pixel interrogation area in steps of 16 pixel are chosen to construct the velocity vector fields. The velocity vector and its components are estimated by taking the mean of all the frames with a proper velocity vector validation. Fig.~\ref{fig:PIV1} shows the velocity vector fields along with the magnitude of the velocity components $v_z$ (Fig.~\ref{fig:PIV1}(a)) and $v_x$ (Fig.~\ref{fig:PIV1}(b)). The magnitude of the velocities are represented by color code of HSV where the blue color corresponds to the minimum value of velocity and red corresponds to the maximum value.  It is clearly seen in the figure that all the particles move from right to left almost in the axial direction as the x-component of the velocity ($v_x$) is almost negligible compared to the axial component of the velocity ($v_z$).  It is also found that the length of the velocity vector field ($v=\sqrt{v_z^2+v_x^2}$) is very small at the right edge (Z-position $\sim 0$ cm) and later gradually increases till $Z \sim 3$ cm and finally it acquires an almost constant length. 
%%%%%%%%%%%%%%%%%%%%%%%%
%%%%%%%%%%%%%%%%%%%%%%%% 
The velocity components ($v_x$, $v_z$) are depicted separately in Fig.~\ref{fig:vxvy}. The solid line represents the variation of $v_z$ whereas the dashed line represents $v_x$ with distance from the right edge. It is clear from the figure that significant contributions to the magnitude of the velocity vector comes from the  $v_z$ component as the particles rarely move in the other direction. It is seen in the experiments that the particles start from their initial velocity and then accelerate towards the port $P_1$ due to neutral streaming and then they attain a terminal velocity. The main force responsible for bringing the dust particles to a terminal velocity is the neutral drag force (an opposition force due to background stationary/moving neutrals) which always acts opposite to the direction of the particle motion. In figure (Fig.~\ref{fig:vxvy}) the solid line shows the particle starting with a finite velocity (at the extreme right edge of the image) and accelerating (up to 2 cm) and finally (beyond 3 cm) achieving an almost constant velocity.        
%%%%%%%%%%%%%%%%%%%%%%%%
%%%%%%%%%%%%%%%%%%%%%%%% 
\subsection{By exciting Dust Acoustic Waves (DAWs) \label{daw}}
\begin{figure}[!ht]
%\centering{\includegraphics[scale=0.60]{../../figures/DAW_with_flow1}}
%\centering{\includegraphics[scale=0.60]{../../figures/fig7_m}}
\centering{\includegraphics[scale=0.60]{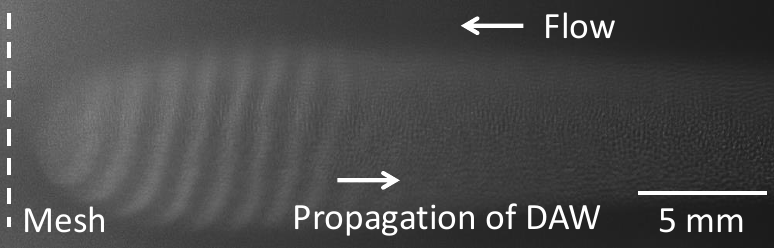}}
\caption{\label{fig:potential} Typical image of propagation of dust acoustic wave (DAW) }
\label{fig:DAW} 
\end{figure}

In this subsection we discuss a new technique to measure the dust fluid velocity that is based on the excitation of a low amplitude Dust Acoustic Wave (DAW) in the medium. This novel technique is applicable even when the particles move with a high velocity and have smaller inter-particle distances as it is independent of the requirement of distinguishing individual particles that is necessary for using the sPIT code method or the PIV analysis. In this technique, after the equilibrium  of dust cloud is achieved by adjusting the pumping and gas feeding rate at lower pressure (P=0.097 mbar) between a mesh and the potential strip, a DAW is excited by applying a sinusoidal voltage ($V_{pp}=100$ volt at a frequency of $f=0.8$ Hz) on a mesh at a discharge voltage of $V_d=350$~volt. The excitation of the Dust Acoustic Waves (DAW) and their subsequent propagation away from the mesh are shown in Fig.~\ref{fig:DAW}. The measured  phase velocity of these waves is around $4-5$~cm/sec and is dependent upon the applied frequency and the plasma and dusty plasma parameters. The experimentally obtained phase velocity is then compared with the theoretical value, $v_{ph}=Z_d\sqrt{\frac{kT_in_d}{m_dn_i}}$ \cite{merlino2012} where, $Z_d$, $n_d$, $n_i$, $kT_i$, $m_d$ are the dust charge number, dust density, ion density, ion temperature and dust mass, respectively. For $Z_d\sim2\times10^4$, $n_d\sim5\times10^{10}/m^3$, $n_i\sim1.2\times10^{15}/m^3 $, $kT_i=0.03$ eV and $m_d=8.7\times10^{-14} kg$ (for a dust grains of radius 2 micron), the phase velocity comes out to be 4.2 cm/sec which agrees well with the experimentally obtained phase velocity. \par  
%%%%%%%%%%%%%%%%%%%%%%%%
\begin{figure}[!ht]
%\centering{\includegraphics[scale=0.75]{../../figures/velocity_flow_rate2_m}}
\centering{\includegraphics[scale=0.75]{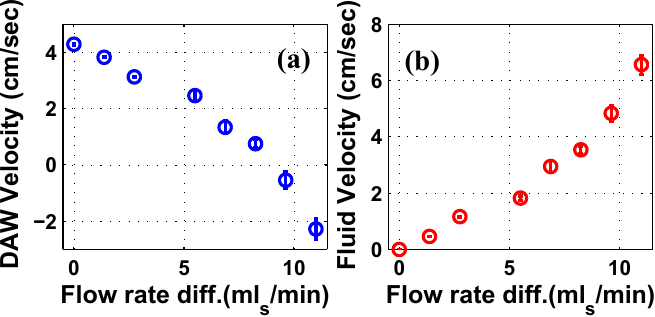}}
\caption{\label{fig:potential}Variation of (a)  phase velocity of dust acoustic wave (DAW)  b) and  fluid flow velocity with the flow rate difference. }
\label{fig:DAW_flow} 
\end{figure}

%%%%%%%%%%%%%%%%%%%%%%%% 
After the excitation of the DAW, the flow of dust is generated with the help of the single gas injection method (discussed in the earlier section \ref{single}) in small steps of the flow rate. As the direction of fluid flow is opposite to the direction of propagation of the DAW, we find that initially the phase velocity of the DAW decreases with the increase of flow rate difference. For a particular higher flow rate difference the DAW becomes almost a standing wave. Then for a further increase in the  flow rate difference, the DAW changes the direction of propagation i.e., it travels in the direction of the neutral flow. The variation of the phase velocity of the DAWs with the flow rate difference is shown in Fig.~\ref{fig:DAW_flow}(a). This figure clearly indicates that the DAW velocity decreases almost linearly with the flow rate change. Subtraction of the original phase velocity ($v_{ph}$) from this velocity (which is the sum of the flow velocity and the $v_{ph}$) directly gives the dust fluid velocity. Fig.~\ref{fig:DAW_flow}(b) shows the dust fluid velocity with the gas flow rate change. However this method also has its disadvantage in that the DAW cannot be excited at higher pressures due to strong damping arising from higher dust neutral collisions. Hence, to measure the flow velocity by this technique, we need to perform the experiments in a pressure range between 0.07~mbar to 0.1~mbar for a particular voltage (V$_d$) $\sim$ 340 volt. \par
%%%%%%%%%%%%%%%%%%%%%%%%
%%%%%%%%%%%%%%%%%%%%%%%% 
Thus the above three techniques to measure the flow velocity, as discussed in this section, have their individual merits and limitations. The sPIT code method is a very efficient and powerful technique to measure the flow velocity as long as one is able to distinguish individual particle motions in the dust component  and the flow is not too high so that the trajectory of each individual particle can be traced in consecutive frames. For a higher density dusty plasma and/or a dust fluid moving with a higher velocity, the sPIT code method fails to measure the velocity accurately. For this reason, it is always recommended to zoom into a small area so that very few number of particles can be tracked for a number of frames. This however limits the information on the spatial variation of flow velocity measurement.  Some of these issues can be addressed  by using the PIV analysis. With the PIV tool one can measure the velocity profile for a reasonably larger field of view. However, this technique also gives better results when in the input images one can clearly identify the motion of individual particles otherwise the measurement of high velocity by this technique becomes difficult. Both these techniques collapse when the fluid velocity turns out to be high and/or the density of the flowing particle is too high so that the particles become indistinguishable. Their efficacy depends on using very high resolution cameras along with high frame rates. In the absence of such a facility, the 3rd technique associated with the excitation of DAWs assumes significance. In this procedure, the DAW acts as a diagnostic tool to provide the information of the fluid velocity albeit with the  limitation that its application is limited to lower pressure discharges (less than 0.12 mbar for the present experiments). Hence an accurate measurement of the flow velocity at higher pressures still remains a challenge and an open issue that needs to be addressed.  
\section{Results and Discussion \label{result}}
In this section, we compare some of the results that we have obtained by employing different techniques to measure the terminal velocity. 
%The terminal velocity is a velocity by which the particles move with a constant velocity. 
In the present set of experiments the dust particles attain a flow velocity due to their interaction with flowing neutrals. The force associated with the momentum transfer from the neutrals to the dust particles is given as $F_n=-m_n\nu_{dn}(v_d-v_n)$ where m$_n$ is the mass of the neutrals, $\nu_{dn}$ is the dust neutral collision frequency and $v_d-v_n$ is the relative velocity of the dust particles with respect to the velocity of the neutrals. Initially, the dust cloud is nearly stationary except for small random velocities associated with the thermal energy of the dust particles. When neutrals with very high velocity are then introduced in the system their collisional impact with the dust particles impart the latter with an average unidirectional momentum that makes them move in the forward direction along the neutral flow. This momentum transfer diminishes as the relative velocity between the two species decreases ultimately leading to a terminal velocity for the dust particles.  In the single (Sec.~\ref{single}) and dual (Sec.~\ref{dual}) gas injection techniques the neutral streaming always carry the particles in its direction whereas in the 3rd technique (Sec.~\ref{confinement}), the flow is generated due to a sudden alteration in the confining potential and there are no streaming neutrals. But in all these three processes, the opposing force comes from a background of moving/stationary neutrals and as a result the dust particles ultimately attain a terminal velocity \cite{thoma2005}. In our experiments, it is found that depending on the plasma and discharge parameters, the particles achieve terminal velocity after travelling a maximum distance of about 2 cms.  \par
%%%%%%%%%%%%%%%%%%%%%%%%
\begin{figure}[!ht]
%\centering{\includegraphics[scale=0.75]{../../figures/velocity_spit_piv_combined}}
\centering{\includegraphics[scale=0.75]{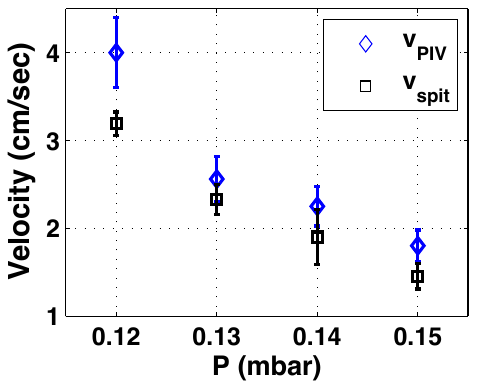}}
\caption{Variation of fluid flow velocity with neutral pressure. Blue diamond corresponds velocity measurement by sPIT code and black square corresponds the velocity measurement by PIV tools.}
\label{fig:PIV_sPIT} 
\end{figure}
%%%%%%%%%%%%%%%%%%%%%%%% 
Fig.~\ref{fig:PIV_sPIT} shows the variation of the terminal velocities with the background neutral gas pressure when the other discharge parameters are kept constant. Data point with \lq \scalebox{0.75}{$\square$}' represents the sPIT data whereas the  \lq $\diamond$' represents the PIV data. It is to be noted that the terminal velocity estimated by both the techniques decreases with the increase of gas pressure. Both the analyses of estimating the flow velocity give almost the same value except when the particle velocity increases (see the data points for $P=0.12$~mbar). As we have discussed, the sPIT code becomes inefficient when the particle velocity becomes higher and this could be the reason why it gives a lower value of the terminal velocity at $P=0.12$ mbar as compared to the PIV analysis. For other pressure regimes the measurements using both the techniques give almost comparable values, although PIV data always gives a slightly higher value of the terminal velocity compared to the values analysed using sPIT code. \par  
%%%%%%%%%%%%%%%%%%%%%%%%
\begin{figure}[!ht]
%\centering{\includegraphics[scale=0.75]{../../figures/velocity_vs_time3_spit_code_m}}
%\centering{\includegraphics[scale=0.75]{../../figures/velocity_with_time}}
\centering{\includegraphics[scale=0.75]{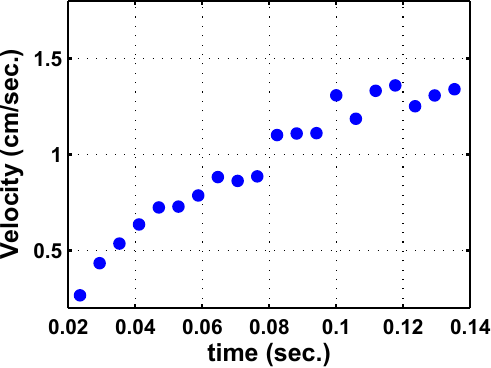}}
\caption{Change of the flow velocity with time for pressure (P) $=0.13$ mbar, discharge voltage (V$_d=310$ volt) and for the flow rate difference of 2.75 ml$_s$/min.}
\label{fig:drag_force} 
\end{figure}
%%%%%%%%%%%%%%%%%%%%%%%% 
We now estimate the forces that are acting on the flowing dust particle using the measurement of the fluid velocity. In the first two experiments, the basic mechanism responsible for generating the dust fluid flow is the neutral gas streaming which always carries the particles along its direction. From the sPIT data analysis it is shown in Fig.~\ref{fig:drag_force}, that the particles are initially accelerated towards the port $P_1$ from a steady state equilibrium position. After travelling a distance of less than about 1 mm, almost all the particles are found to attain a terminal velocity within about 100~msec due to the resultant opposing forces acting on them. The resultant force acting on the particles could be attributed due to electrostatic force, ion drag force or/and neutral drag force. In present set of experiments, the discharge is operated at a pressure where the neutral density lies between $10^5$ to $10^6$ times higher than the ion density. Hence, the ion drag force acting on the dust grains is considerably small compared to the neutral drag force  \cite{jaiswal2015}. From a detailed study on the variation of plasma and floating potentials at different discharge conditions \cite{jaiswal2015}, it is found that the axial component of electric field is very small and therefore the contribution of electrostatic force on the flow of dust particles can also be neglected. To acertain this, an experiment is also performed to measure the flow velocity at different discharge voltages. It is observed that the particles changes only their equilibrium height with the voltages but the flow velocity remains unaffected.  At the time when the particles acquire terminal velocity ($v_r$), the neutral drag force can be expressed as $F_{ND}=-\frac{4}{3}\gamma_{Eps}{\pi}r_d^2m_nN_nv_{thn}v_r$ \cite{thomas2005}, where, $m_n$, $N_n$, $v_{thn}$ and $v_n$ are the mass, background density, thermal and drift velocities of the neutrals respectively. $\gamma_{Eps}$ represents the Epstein drag coefficient which varies from $1$ to $1.4$ depending upon the types of reflection \cite{epstein1924}. Therefore, the exact value of $\gamma_{Eps}$ is essentially needed to find out the neutral drag force for our experiments. For the estimation of $\gamma_{Eps}$, we then calculate the slope ($\frac{dv}{dt}$) of the straight line of Fig.~\ref{fig:drag_force} till 80 msec and multiply it by the average mass of the dust particles ($m_d=6.2\times10^{-13}$ kg for the dust particles of radius r$_d=4.59 \times10^6~\mu$m) to get the accelerating force which is later equated with $F_{ND}$. For a given value of terminal velocity (in this present case it is $\sim 1.2$ cm/sec) and plasma/dusty plasma parameters such as $m_n=6.68\times10^{-26} kg$, $N_n=2.7\times10^{21}$/m$^3$, $v_{thn}=427.7$ m/sec, $\gamma_{Eps}$ comes out to be $\sim~1.07$. 
With the help of $\gamma_{Eps}$ and for a wide range of terminal velocities as shown in  Fig.~\ref{fig:PIV_sPIT} the neutral drag force is estimated to be $1\times 10^{-13}$N to $3\times 10^{-13}$N. \par
%%%%%%%%%%%%%%
\section{Conclusion \label{conclusion}}
To conclude, in this paper, we have presented and discussed a variety of experimental means of generating and measuring flows in a dusty plasma fluid. The techniques have been tested in a series of experiments carried out in the DPEx device for a dusty plasma of MF/kaolin particles embedded in a DC glow discharge Ar plasma. The initial steady state equilibrium dust cloud formed in a confining potential well by adjusting the pumping speed and the gas flow rate can be made to flow by using streaming neutrals introduced from single or duel gas injection ports or by suddenly lowering the confining potential. The resultant dust fluid velocity can be measured by using a sPIT code or a PIV analysis. Another novel way is by exciting DAWs - a technique that we have successfully tried out for the first time in our experiments. Each method has its strengths and limitations which we have pointed out on the basis of our experimental findings. We have also provided estimates of the terminal velocities that the dust component can acquire based on a theoretical evaluation of the neutral drag force that acts on the dust. We believe our findings can be usefully employed to facilitate future experimental explorations of linear/nonlinear wave excitations and other phenomena associated with flowing complex plasmas such as their interactions with charged objects or potential barriers.  
%%%%%%%%%%%%%      
\section{References.\label{bibby}}

\end{document}